\documentclass[12pt]{article}
\usepackage{epsfig, amssymb}
\newcommand{\be}{\begin{equation}}
\newcommand{\ee}{\end{equation}}
\newcommand{\bea}{\begin{eqnarray}}
\newcommand{\eea}{\end{eqnarray}}
\newcommand{\bA}{\begin{array}}
\newcommand{\eA}{\end{array}}
\newcommand{\bc}{\begin{center}}
\newcommand{\ec}{\end{center}}
\newcommand{\al}{\alpha}

\newcommand{\ra}{\rightarrow}
\newcommand{\del}{\partial}
\newcommand{\ie}{{\it i.e.}}
\newcommand{\eg}{{\it e.g.}}

\newcommand{\Nf}{${\cal N}{=}4$}
\newcommand{\Nt}{${\cal N}{=}2$}
\newcommand{\No}{${\cal N}{=}1$}

\def\BC{{\Bbb C}}
\def\BR{{\Bbb R}}
\def\BZ{{\Bbb Z}}

\begin{document}

\begin{titlepage}

\bc

\hfill  {Duke-CGTP-03-03} \\
\hfill  {NSF-KITP-03-77} \\
\hfill  {\tt hep-th/0309171} \\
        [22mm]

{\Huge Coarse-graining quivers}\\ 
\vspace{10mm}

{\large K.~Narayan and M.~Ronen Plesser} \\
\vspace{3mm}
{\small \it Center for Geometry and Theoretical Physics, \\}
{\small \it Duke University, \\}
{\small \it Durham, NC 27708.\\}
\vspace{1mm}
{\small Email : narayan, plesser@cgtp.duke.edu}\\

\ec
\medskip
\vspace{35mm}

\begin{abstract}
We describe a block-spin-like transformation on a simplified subset of 
the space of supersymmetric quiver gauge theories that arise on the 
worldvolumes of D-brane probes of orbifold geometries, by sequentially 
Higgsing the gauge symmetry in these theories. This process flows to 
lower worldvolume energies in the regions of the orbifold moduli space 
where the closed string blowup modes, and therefore the expectation 
values of the bifundamental scalars, exhibit a hierarchy of scales. 
Lifting to the ``upstairs'' matrices of the image branes makes contact 
with the matrix coarse-graining defined in a previous paper. We describe
the structure of flows we observe under this process. The quiver 
lattice for $\BC^2/\BZ_N$ in this region of moduli space deconstructs 
an inhomogeneous, fractal-like extra dimension, in terms of which our 
construction describes a coarse-graining of the deconstruction lattice.

\end{abstract}

\end{titlepage}

\newpage 
\begin{tableofcontents}
\end{tableofcontents}

\vspace{5mm}

\section{Introduction}

The past few years have seen the emergence of deep interrelations
between gauge theory and gravity. In particular, Matrix theory
\cite{BFSS} and more generally, the gauge/gravity correspondence
\cite{malda9905} show that string/M theories in certain backgrounds
are equivalent to gauge theories. In the context of these recent
advances, an effective description in terms of a smooth spacetime
emerges in an appropriate large-$N$ limit from the dynamics of matrix
entries.  In general, the worldvolume scalars describing the motion of
branes in the transverse space are matrices. Thus various
configurations of collections of D-branes are described by matrix
representations.

The emergence of a continuum spacetime description from a discrete
matrix index is reminiscent of the emergence of a continuum limit in
lattice theories of critical phenomena.  In these cases, the 
renormalization group provides a useful conceptual and technical
framework for understanding the continuum limit and studying its
properties.  The continuum limit emerges in an RG approach as a fixed
point of a transformation in which the dynamical degrees of freedom are
``thinned'' by removing some high-energy modes.

In \cite{kn0211} an attempt was made to apply a suggestive analogy to
understand the interrelations between smooth spacetime and matrix 
representations of D-brane configurations.  The matrices representing 
configurations of an infinite collection of branes were subjected to a 
block-spin-like transformation in which nearest-index blocks were 
averaged.  The observation of \cite{kn0211} is that under this 
transformation the matrices become ``less off-diagonal'' in the sense 
that if initially the matrix entries satisfied 
$ X_{ab}\sim X_{aa} q^{|b-a|}$ for some $q < 1$, then the resulting 
block matrices satisfy the same condition with a parameter 
${\tilde q} \sim q^2$.  

It is tempting to think of this transformation as a coarse-graining of
the space in which the branes move.  A precise formulation is however
difficult, among other things because the properties of the matrices
used to make the construction are not invariant under the gauge
symmetry of the problem, and hence cannot be formulated in physical
terms.  In this work, we will try to realize a similar idea in a slightly
different context.  We consider the worldvolume dynamics of a D-brane
in the vicinity of a quotient singularity $\BC^r/\BZ_N$ in the
transverse space, selected so that the resulting low-energy gauge
theory is supersymmetric.  

The low-energy dynamics on the worldvolume of a D3-brane at a quotient
singularity $\BC^3/\BZ_N$ is well-known \cite{douglasmoore}.  It is
obtained from the \Nf\ theory with gauge group $U(N)$ describing
the motion of $N$ ``image'' D3-branes on the covering space $\BC^3$ by
a projection to $\BZ_N$ invariant degrees of freedom.  The group
action on the transverse space must be augmented by an action on the
Chan-Paton indices to determine a projection.  Different choices of
this representation correspond to the various three-dimensional BPS
branes.  In addition to the D3-brane we are after, these include
D5-branes wrapping two-cycles localized near the singularity and
D7-branes wrapping such four-cycles.  These states are thus ``pinned''
to the singular locus. This correspondence between representations of
the discrete group and the homology of the quotient space is the
(generalized) McKay correspondence \cite{mckay}.  In our case we will use
the regular representation, corresponding to a D3-brane free to move
on the quotient space.

The worldvolume dynamics that emerges from the projection is governed
by a $U(1)^N$ gauge group, with chiral multiplets transforming in
various bifundamental representations.  In general there are also
neutral ``adjoint'' multiplets.   The interactions are encoded in a
cubic superpotential that is simply the restriction of the cubic
superpotential of the \Nf\ theory to the fields surviving the
projection.

The closed string background determines the parameters of the
worldvolume action.  In particular, the twisted sector massless fields
whose expectation values label the moduli space of blowups of the
quotient singularity appear as Fayet-Iliopoulos D-terms in the
worldvolume action.  For each choice of these, the theory has a moduli
space of classical vacua which is simply the partially resolved
quotient space.  At generic points in the moduli space the gauge
symmetry is broken to $U(1)$, but by a suitable choice of blowup
parameters, essentially corresponding to a sequence of partial
resolutions with widely different sizes for the exceptional cycles, we
can set up a hierarchy of scales such that the symmetry breaks
sequentially, \eg\ for $N=2^k M$ we can find $U(1)^{2^k M}\to
U(1)^{2^{k-1}M}\to\cdots\to U(1)^M\cdots$.

The ``thinning'' of worldvolume degrees of freedom we observe here is
of course nothing mysterious.  We are simply observing the Higgs
mechanism for carefully chosen Higgs expectation values.  One can
however attempt to make a connection to the ideas discussed 
above\footnote{In particular, see the discussion on branes arranged in 
quasi-linear chains in Appendix A of \cite{kn0211}.} along the following 
lines : In the limit of large $N$, there is a (distinct) region in the 
moduli space in which the low-energy theory effectively lives on the 
discretization of the circle approximated by the quiver lattice with $N$
sites.  A transformation that halved $N$ in this ``deconstruction'' region
would be naturally thought of as a coarse-graining of this lattice.  Of 
course, our construction works in a different region of moduli space, but 
we expect that the qualitative structure, \ie\ the directions of the 
flows and the fixed points, are the same as what would be obtained via 
conventional real space renormalization in the extra dimensions.

Another heuristic connection to the ideas above is obtained by lifting 
the Higgsing process we observe to the ``upstairs'' $N\times N$ matrices 
of the $N$ image branes. The reflection of the Higgsing process in 
these matrices coincides with the matrix coarse-graining defined in 
\cite{kn0211}, thus providing a concrete field theoretic realization of
that matrix coarse-graining.

Some words on organization : in section 2, we study the coarse-graining 
of the $\BC^2/\BZ_N$ quiver via sequential Higgsing and the more general 
$\BC^3/\BZ_N$ quivers with action $(1,a,-a-1)$ in section 3 and the 
corresponding flows. In section 4, we lift this Wilsonian probe 
worldvolume RG to the ``upstairs'' $N\times N$ matrices, thereby making 
contact with the matrix coarse-graining of \cite{kn0211}. In section 5, 
we map out the structure of the flows we observe in the space of these 
supersymmetric quivers and finally end with some conclusions in section 6.

\section{$\BC^2/\BZ_N$ quotient}

Consider a D3-brane probe near the tip of a $\BC^2/\BZ_N$ orbifold.
We choose our coordinates so that the 
brane worldvolume coordinates are $0123$ while the remaining six
transverse coordinates are reorganized into complex coordinates 
$X=z^4+iz^5,Y=z^6+iz^7,Z=z^8+iz^9$. We take the orbifold action to be 
\be\label{geomactionC2}
(X,Y,Z) \ra (\omega X, \omega^{-1} Y, Z), 
\qquad \qquad \qquad \omega=e^{2\pi i/N}\ .
\ee

\begin{figure}
\bc
\epsfig{file=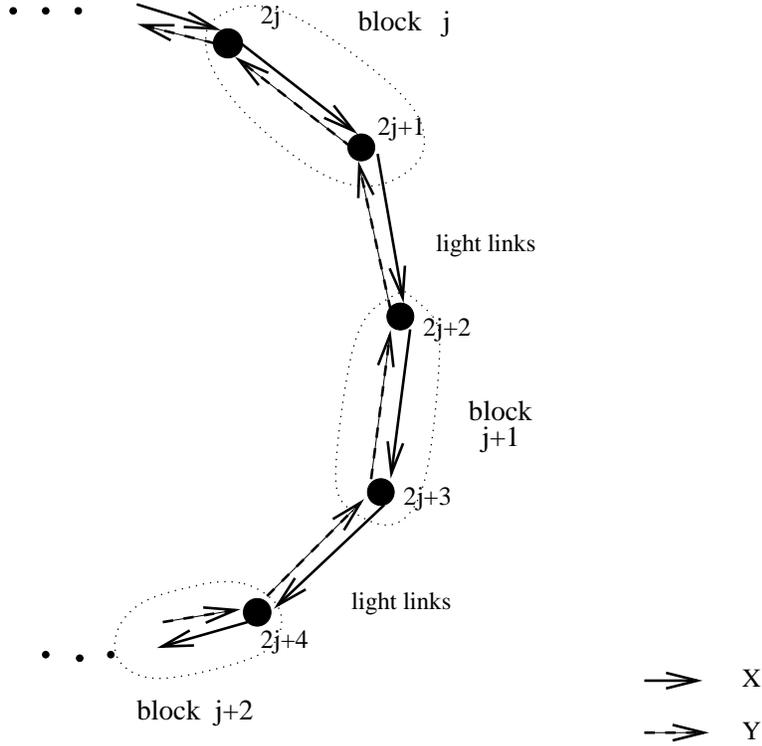, width=10cm}
\caption{$\BC^2/\BZ_{2M} \ra \BC^2/\BZ_{M}$ : Coarse-graining as 
$(2j,2j+1)\to j$ by condensing the link field $X_{2j,2j+1}$. The quiver 
for the residual light link fields represents $\BC^2/\BZ_{M}$.}
\label{fig2}
\ec
\end{figure}

Implementing the construction of \cite{douglasmoore} (see also \eg\
\cite{joep9606} \cite{jm9610}) we find that the components of the $U(N)$ 
gauge fields on the covering space which survive the projection 
correspond to a $U(1)^N$ gauge symmetry, with vector multiplets we denote 
by $A^{\mu}_i$.  The components of the chiral multiplets associated to 
the transverse coordinates which survive the projection we denote as 
$Z_{ii}, X_{i,i+1}, {\bar X}_{i+1,i}, Y_{i+1,i}$ and ${\bar Y}_{i,i+1}$, 
where $X_{i,i+1}$ carries charge $\pm 1$ under gauge groups $U(1)_i$ and 
$U(1)_{i+1}$ respectively, while $Y_{i+1,i}$ carries charge $\mp 1$ under 
the same gauge groups. $Z_{ii}$ are neutral fields.  The quotient theory
preserves \Nt\ supersymmetry.  
This theory can be represented by the quiver diagram depicted in 
figure (\ref{fig2}).

For ease of notation, let us relabel 
$X_{i,i+1} \ra X_i$ ($i$ being the node from which the arrow emanates) 
and $Y_{i+1,i} \ra Y_i$ ($i$ here being the node at which the arrow ends). 
The superpotential of the quotient theory is given by the truncation of 
the cubic \Nf\ superpotential to the surviving fields
\bea\label{Ntsuperpot}
W &=& {\rm Tr}\ X[Y,Z] \biggr|_{projected} = 
\sum_{i=1}^N \biggl( X_{i,i+1} Y_{i+1,i} Z_{i,i} - X_{i-1,i} Z_{i,i} 
Y_{i,i-1} \biggr) \nonumber\\
&\equiv& \sum_{i=1}^N Z_i (X_i Y_i - X_{i-1} Y_{i-1})
= \sum_{i=1}^N X_i Y_i (Z_i - Z_{i+1})\ .
\eea
The overall normalization is unimportant for our analysis here. The 
F-term equations resulting from this superpotential are
$\del W/ \del \Psi_k=0$ for all fields $\Psi_k$, giving for the vacuum
values of the fields 
\be\label{NtFterms}
X_i Y_i = X_{i-1} Y_{i-1}, \qquad Z_i Y_i = Z_{i+1} Y_i, \qquad 
Z_i X_i = Z_{i+1} X_i
\ee
while the D-term equations for the expectation values for gauge group 
$U(1)_i$ are 
\be\label{NtDterms}
D_i
= |X_i|^2 - |Y_i|^2 - |X_{i-1}|^2 + |Y_{i-1}|^2 = \rho_i
\ee
where the $N$ real FI D-terms $\rho_i$ are determined as noted above
by the expectation values of twisted sector closed string fields.
Note that the form of (\ref{NtDterms}) implies that $\sum_i \rho_i =
0$, leaving $N-1$ independent real FI parameters as expected,
corresponding to the $N-1$ twisted sectors (and independent blowup
parameters) of the singularity.

At $\rho_i=0$, the moduli space of classical vacua can be
parameterized by the gauge invariant polynomials in the chiral
multiplets, subject to the relations (\ref{NtFterms}).  
The invariant polynomials are generated by 
\be
C_i=(X_i Y_i),  \qquad A=(X_1 X_2 \ldots X_N), \qquad 
B=(Y_1 Y_2 \ldots Y_N)\ .
\ee
Then it is clear from (\ref{NtFterms}) that $C_i=C_{i-1}=C$, and
$A B = C^N$.  The moduli space is thus precisely the 
$\BC^2/\BZ_{N}$ orbifold singularity which we here realize as a
hypersurface in $\BC^3$.

\subsection{Sequential Higgsing}

Deforming the closed string background to resolve the orbifold forces
the chiral multiplets to acquire nonzero expectation values,
spontaneously breaking the $U(1)^N$ gauge symmetry.  Generically the
symmetry is broken, leaving only the decoupled overall $U(1)$ symmetry
and an \Nf\ theory at low energies.  In this section we will show that
for suitably chosen blowup parameters the pattern of breaking is
sequential with the rank of the unbroken gauge group reduced at each
of many scales.  For simplicity and concreteness, we will halve the
rank at each scale.  

Let us thus assume that $N=2M$ is even.
To begin with, consider the blowup pattern that resolves the $A_{N-1}$
quotient singularity to an $A_{M-1}$ quotient.  In terms of the FI
D-terms this corresponds to $D_i=\rho_i = (-1)^i\rho$.  The maximally
symmetric vacuum with these parameters is represented by the
expectation values
\be
X_{2j}=\sqrt{\rho}, \qquad X_{2j-1}=0, \qquad Y_i=0=Z_i\ .
\ee
This choice of expectation values breaks each pair (or block) 
$U(1)_{2j}\times U(1)_{2j+1}\to U(1)$, reducing the gauge group in all 
to $U(1)^{M}$.

Inserting these expectation values into the superpotential
(\ref{Ntsuperpot}) we see that some of the $Y$ and $Z$ fields acquire
a mass.  Explicitly, defining
\be
Z_j^+={1\over 2}(Z_{2j}+Z_{2j+1}), \qquad 
Z_j^-={1\over 2}(Z_{2j}-Z_{2j+1}), \qquad 
j=1,2,\ldots M={N\over 2}
\ee
we can write the quadratic part of $W$ as
\be
W_2 = \sqrt{\rho}\sum_{j=1}^M Y_{2j} Z_j^-\ .
\ee
Of course, $X_{2j}$ is eaten by the super-Higgs mechanism, leaving as the
light degrees of freedom $X_{2j+1}, Y_{2j+1}$, and $Z_j^+$.  We see
that their charges under the remaining gauge group together with the
remaining superpotential 
\be
W_{eff} = \sum_{j=1}^{M} X_{2j+1} Y_{2j+1} (Z_j^+ - Z_{j+1}^+)
\equiv \sum_{j=1}^{M} X_{j,j+1}Y_{j+1,j}(Z_{j,j}-Z_{j+1,j+1}),
\ee
(in the last line, we have relabelled the two branes in each block as
$(2j,2j+1) \ra j$) are precisely what our original quotient
construction would produce for the case of a $\BZ_M$ quotient.  This
shows how the partial blowup is realized in the worldvolume Higgs
mechanism. 

We can realize the RG-like ideas mentioned above by repeating the
above process sequentially at a hierarchical sequence of scales.  In
other words, if $M$ is even, we can now repeat the process with $D_j =
(-1)^j \rho'$ for $\rho'\ll\rho$, etc.

\subsection{Some comments on deconstructive RG}

In this section, we draw some relations to coarse-graining the 
deconstructed extra dimension(s). The quiver theory on a D-brane probe 
at a distance $d=R/2\pi$ from the $\BC^2/\BZ_N$ orbifold tip deconstructs 
\cite{decons} a homogeneous extra dimension in a certain field theory 
limit where the quiver lattice is uniform \cite{deconsLST0110}. 
Explicitly, consider giving equal expectation values to the 
bifundamentals (with $Z_i=0$)
\be
X_i=X_{i-1}=h, \qquad Y_i=Y_{i-1}=h\ .
\ee
This solves the F- and D-term equations with vanishing D-term parameters
$D_i=0$. Then the potential energy for fluctuations about such a vacuum is
\be
V \sim \sum_i 2h^2 |Z_i - Z_{i+1}|^2 + \ldots
\ee
which looks like a lattice discretization of the kinetic term for $Z$
in a new continuum dimension -- in the large $N$ limit, we have a sharp
orbifold tip, the continuum approximation is exact and this set of 
expectation values deconstructs a uniform spatial dimension \cite{decons} 
\cite{deconsLST0110}. The gauge symmetry remains $U(1)^N$ all through 
intermediate energy scales and breaks down to the diagonal $U(1)$ in the 
far IR.

It is thus clear that the region of moduli space that we have described
in the previous section is different : the bifundamental expectation 
values reflect the hierarchy of scales present in the closed string 
blowup modes\footnote{It is straightforward to construct a set of Higgs 
expectation values that interpolates between the equal expectation 
values limit and our hierarchical expectation values limit, essentially 
by tweaking the closed string blowup modes $\rho_i$ (see Appendix).}. 
This hierarchy of scales deforms the lattice so as to make every 
alternating set of links (the $X_{2j,2j+1}$s) more massive than the 
others. Continuing this iteratively, we see that the quiver lattice 
generates an inhomogeneous fractal-like extra dimension, in the region 
of moduli space we study. 
The parameters in the field theory in the hierarchical region are
\be
2\pi R=\sum_i a_i, \qquad a_i={1\over g v_i},
\ee
where $v_i = \langle X_i\rangle$. 
Under the process of ($2\times 2$)-block Higgsing, the discretization 
$N$ of the quiver lattice reduces to ${N\over 2}$. Furthermore, since we 
have integrated down to lower scales $v' \ll v$, the accuracy to which 
length scales can be resolved decreases. The smallest spacing between 
the neighbouring lattice nodes changes to 
$a'={1\over g v'} \gg {1\over g v}=a$. Thus this process is naturally 
thought of as coarse-graining the lattice along the deconstructed 
dimension. Now if our hierarchies are arranged so that 
$a'=2a,\ a''=2a'=2^2a,\ldots$, then the smallest lattice spacing 
effectively doubles at every step of our Higgsing. Thus the lattice 
coarsens, as is expected for a real space RG-like transformation in the 
extra dimension. 

Indeed due to the high amount of supersymmetry, perhaps our Higgsing
calculations closely approximate a conventional block-spin-type 
renormalization in the extra dimension, besides qualitative matches. 
It would be interesting to explicitly study a modification of our 
calculations which details the precise relation between our calculations 
here and RG in the usual deconstruction region of moduli space, perhaps
using the interpolating expectation values described in the Appendix.

\section{$\BC^3/\BZ_N$ quotient}

Consider a single D3-brane probe near the tip of a $\BC^3/\BZ_N$ orbifold 
\cite{mrdbrgdrm} \cite{drmplesserHorizons} with the orbifold geometric 
action\footnote{In general, with an orbifold action $(a_1,a_2,a_3)$ 
given as $\BZ^I\ra \omega^{a_I} \BZ^I$, some fraction of the \Nf\ 
supersymmetry is preserved if the orbifold action lies within $SU(3)$ 
(instead of $U(3)$) which requires $\sum_I a_I=0({\rm mod} N)$.
If one of the $a_I$s is zero, then the $Z_N$ lies in an $SU(2)$ 
subgroup of the $SU(3)$ thus preserving \Nt\ supersymmetry (eight 
supercharges) as in the $\BC^2/\BZ_N$ case, while if all of the $a_I$ are 
nonzero, then we preserve \No\ supersymmetry in four dimensions. The 
$(n,-n,0)({\rm mod} N)$ case is equivalent to $(1,-1,0)$ (since we can 
relabel the quiver nodes) so that there is only way to realize an \Nt\ 
theory while there are several ways to realize \No. More generally, 
taking $\sum_I a_I\neq 0({\rm mod} N)$ breaks all supersymmetry.} 
$(a_1,a_2,a_3)$, subject to $\sum_ia_i=0({\rm mod}N)$, given by
\be\label{geomactionC3a1a2a3}
(X,Y,Z) \ra (\omega^{a_1} X, \omega^{a_2} Y, \omega^{a_3}Z)
\qquad \qquad \qquad \omega=e^{2\pi i/N}\ .
\ee
Then the \No\ fields surviving the projection 
are $A^{\mu}_{i},X_{i,i+a_1},{\bar X}_{i+a_1,i}, Y_{i,i+a_2},
{\bar Y}_{i+a_2,i}$, $Z_{i+a_3,i}$ and ${\bar Z}_{i,i+a_3}$. The surviving
gauge group is thus $U(1)^N$. The $X_{i,i+a_1}$ carries charge $\pm 1$ 
under gauge groups $U(1)_i$ and $U(1)_{i+a_1}$, $Y_{i,i+a_2}$ carries 
charge $\pm 1$ under gauge groups $U(1)_i$ and $U(1)_{i+a_2}$ while 
$Z_{i+a_3,i}$ carries charge $\mp 1$ under $U(1)_i$ and $U(1)_{i+a_3}$. 
In figure~\ref{fig4}, figure~\ref{fig5} and figure~\ref{fig6}, we have 
shown the quivers for some specific cases. As before, for ease of 
notation, let us relabel $X_{i,i+a_1}\ra X_i,Y_{i,i+a_2}\ra Y_i$ ($i$ 
being the node from which the arrow emanates), and $Z_{i+a_3,i}\ra 
Z_{i+a_3}$ ($i+a_3$ here being the node at which the arrow emanates). 
Then the orbifolded theory preserves \No\ supersymmetry
and has the superpotential
\bea\label{N1superpota1a2a3}
W &=& {\rm Tr}\ X[Y,Z] \biggr|_{projected} = 
\sum_{i=1}^N (X_{i,i+a_1}Y_{i+a_1,i+a_1+a_2} - Y_{i,i+a_2} 
X_{i+a_2,i+a_2+a_1}) Z_{i+a_1+a_2,i} 
\nonumber\\
&\equiv& \sum_{i=1}^N Z_{i+a_1+a_2} (X_i Y_{i+a_1} - Y_i X_{i+a_2})
= \sum_i X_i (Y_{i+a_1} Z_{i+a_1+a_2} - Y_{i-a_2} Z_{i+a_1})\ .
\eea
The F-term equations then are $\del W/\del \Psi_k = 0$ for all fields 
$\Psi_k$ while the D-term equations for the gauge group $U(1)_i$ are 
\be\label{N1Dterms}
D_i = |X_i|^2 + |Y_i|^2 - |Z_{i+a_3}|^2 - |X_{i-1}|^2 - |Y_{i-1}|^2 
+ |Z_{i-a_3}|^2 = \rho_i\ .
\ee

There is a well-known subtlety we should mention here:  in general the
gauge theories we obtain in this way are anomalous (this problem as
well as its resolution was pointed out in \cite{douglasmoore} and
addressed in detail in \cite{drmplesserHorizons,uranga9808}), and cannot
describe a consistent worldvolume dynamics.  The anomalies are
cancelled by anomalous transformation properties of twisted sector
closed-string fields (which do not of course appear in our worldvolume
Lagrangians).  A Dine-Seiberg like mechanism in fact {\it breaks\/}
the gauge symmetry.  

For our discussion, the main impact of this is that the D-term
parameters $\rho_i$ should now be considered not fixed background
parameters but rather fluctuating dynamical degrees of freedom.  A
physical way to recognize the difference is that in the
codimension-four case of the previous section, we could fix the blowup
parameters by a measurement far from the brane worldvolume in the $Z$
directions (or, by setting suitable boundary conditions there, fix
them).  In the present case the blowup modes are pinned to the
codimension-six singular locus, and cannot be fixed far from the
brane.  

For the classical discussion we present, this is not crucial.  We will
still find a region in the moduli space of vacua where the pattern of
symmetry breaking follows a hierarchical pattern as in the previous
section, and where the theories at intermediate scales are essentially
quotient theories with smaller values of $N$.  Quantum mechanically,
however, the significance of these calculations is less clear.  It is
possible, for example, that a potential on the moduli space is
generated and lifts the region in question.  Since we are 
considering the D-brane probe worldvolume theory at weak coupling 
$g_s\sim 0$ with $\al'\ra 0$ in the substringy regime near the orbifold
tip (\ie\ the expectation values $v$ of worldvolume scalars satisfy 
$v \ll l_s^{-1}$), we hope that quantum corrections do not invalidate 
our calculations here.

\subsection{Sequential Higgsing : $(1,2a+1,-2a-2) \ra (1,a,-a-1)$}

In this subsection, we specialize to the case $(a_1,a_2,a_3) =
(1,2a+1,-2a-2)$. 
Consider the D-term configuration $D_i = (-1)^i\rho$.  The maximally
symmetric vacua are then represented by
\be
X_{2j}=\sqrt{\rho}, \qquad j=1,\ldots,M={N\over 2}
\ee
as the only nonzero expectation values.  This Higgses the 
$U(1)\times U(1)$ in each block down to the diagonal $U(1)$ as above. 
Writing the superpotential as a quadratic piece $W_2$ containing the 
Higgs expectation value for $X_{2j}$ -- these give rise to mass terms 
in the effective potential -- plus a cubic piece $W_3$ gives
\bea
&& W=W_2+W_3,\nonumber\\
&& {} 
W_2=\sum_{j=1}^M X_{2j} (Y_{2j+1} Z_{2j+2a+2} - Y_{2j-2a-1} Z_{2j+1})
\nonumber\\
&& {} \qquad
= \sqrt{\rho} \sum_j Y_{2j+1} (Z_{2j+2a+2} - Z_{2j+2a+3}) 
= \sqrt{\rho} \sum_j Y_{2j+1} Z_j^-,
\nonumber\\
&& {} 
W_3=\sum_{j=1}^M X_{2j+1} (Y_{2j+2} Z_{2j+2a+3} - Y_{2j-2a} Z_{2j+2})
\eea
where we have defined $Z_j^{\pm}={1\over 2}(Z_{2j}\pm Z_{2j+1})$. Then
the $Y_{2j+1}$ and $Z_j^-$ fields appearing in $W_2$ become massive. The
classical equations of motion for $Y_{2j+1}$ following from $W$ give
\be
{\del W \over \del Y_{2j+1}}={\del W_2 \over \del Y_{2j+1}}=0=Z_j^-
={1\over 2}(Z_{2j}-Z_{2j+1})
\ee
so that
\be
Z_j^+={1\over 2}(Z_{2j}+Z_{2j+1})=Z_{2j}=Z_{2j+1}
\ee
\begin{figure}
\bc
\epsfig{file=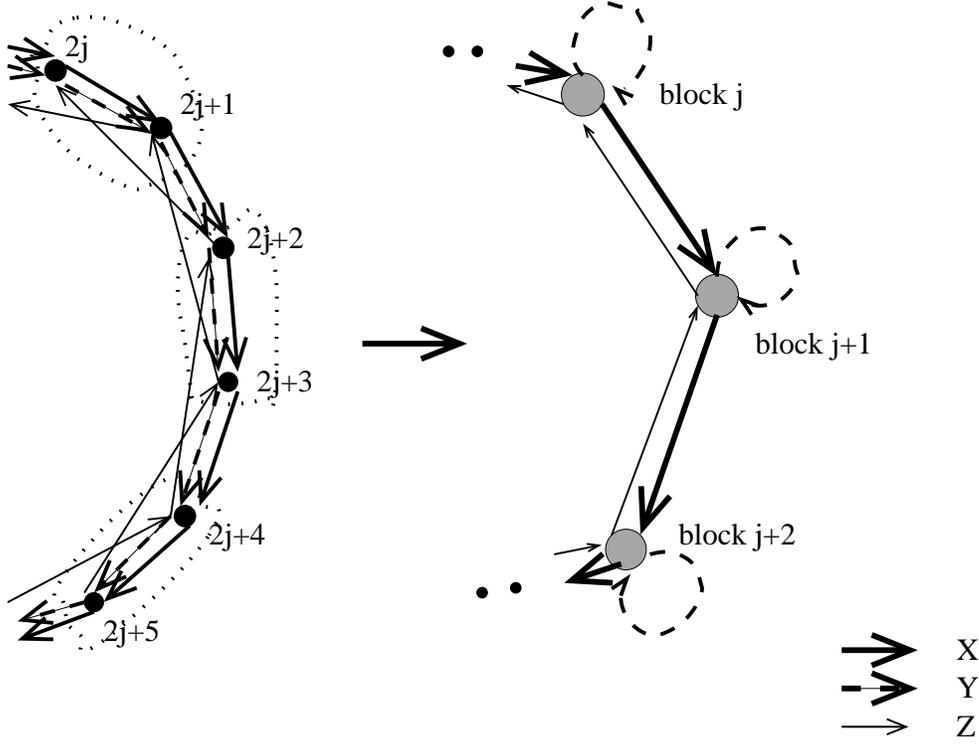, width=13cm}
\caption{$\BC^3/\BZ_N (1,1,-2) \ra \BC^2/\BZ_{N/2} (1,0,-1)$ : 
Coarse-graining as $(2j,2j+1)\ra j$  by condensing the $X_{2j,2j+1}$ link. 
The $Y_{2j,2j+1}$ links become light neutral fields while $X_{2j-1,2j}$ and 
$Z_j^+={1\over 2}(Z_{2j,2j-2}+Z_{2j+1,2j-1})$ become light bifundamentals.
This new quiver now represents $\BC^2/\BZ_{N/2}$. We have not shown the 
$Y_{2j-1,2j}$ and $Z_j^-$ links that become massive under the Higgsing.}
\label{fig4}
\ec
\end{figure}
The cubic part of the superpotential can be rewritten as
\bea
&& {} W_3=\sum_{j=1}^M X_{2j+1} [Y_{2j+2} Z_{j+a+1}^+ - Y_{2j-2a} 
Z_{j+1}^+] 
=\sum_{j=1}^M Z_j^+ (X_{2j-2a-1} Y_{2j-2a} - X_{2j-1} Y_{2j-2a-2})
\nonumber\\
&& {}\qquad \equiv \sum_{j=1}^{M} 
Z_{j,j-a-1} (X_{j-a-1,j-a} Y_{j-a,j} - X_{j-1,j} Y_{j-a-1,j-1})
\eea
In the last line, we have as before relabelled $(2j,2j+1)\ra j$, so that 
this now looks like an \No\ superpotential for the light fields 
$X_{2j+1}, Z_j^+, Y_{2j}$ relabelled as 
$X_{j,j+1},Z_{j,j-a-1},Y_{j,j+a},\ j=1,\ldots,{M}$. This is succinctly 
expressed diagrammatically as shown in some specific examples in 
figure~\ref{fig4}, figure~\ref{fig5} and figure~\ref{fig6}. More 
generally, looking at the quiver for $(1,2a+1,-2a-2)$, it is clear that,
\eg\ the light link $Y_{2j,2j+2a+1}$ connects the $j$-th and $(j+a)$-th 
blocks, so that the new quiver is $(1,a,-a-1)$. Thus the residual quiver 
represents a $\BC^3/\BZ_{M}$ geometry with the orbifold action $(1,a,-a-1)$
on the coordinates $X,Z^+,Y$.
\begin{figure}
\bc
\epsfig{file=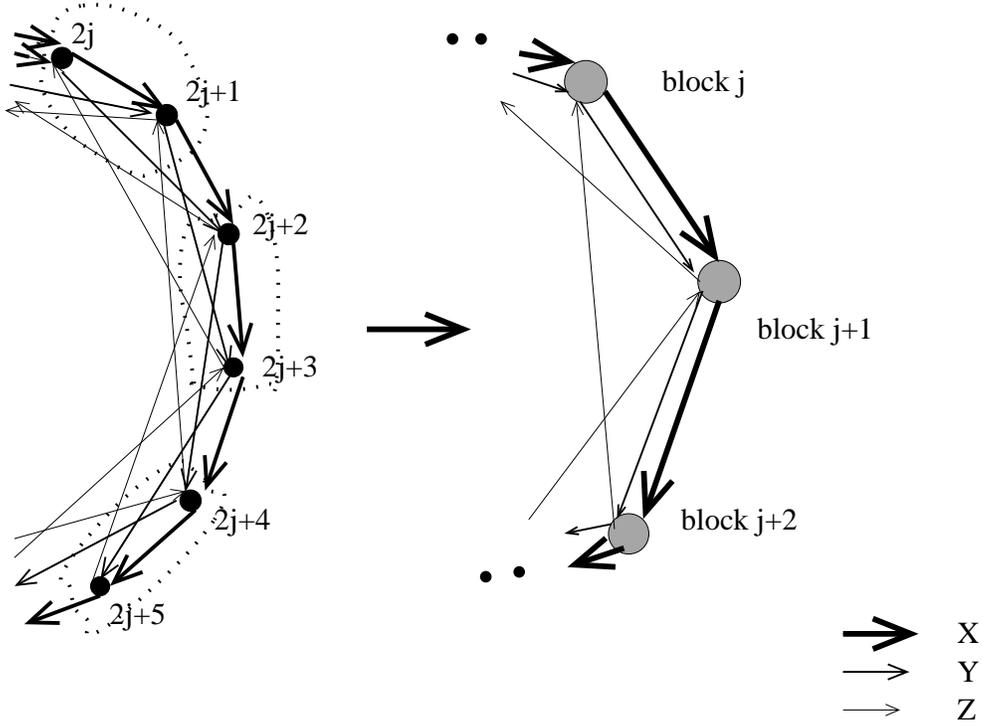, width=13cm}
\caption{$\BC^3/\BZ_N : (1,2,-3) \ra (1,1,-2)$ : Coarse-graining as 
$(2j,2j+1)\ra j$ by condensing the $X_{2j,2j+1}$ link. Two of the four
effective links between, \eg\ blocks $j$ and $j+1$ become massive --
these represent $Y_j^-, Z_{2j+3,2j}$. Retaining only the light link 
fields gives the quiver on the right, \ie\ $(1,1,-2)$.}
\label{fig5}
\ec
\end{figure}

\subsection{Sequential Higgsing : $(1,2a,-2a-1) \ra (1,a,-a-1)$}

This subsection, very similar to the previous one, specializes to the case
$(1,2a,-2a-1)$. Consider the D-term configuration $D_i = (-1)^i\rho$.
The maximally symmetric vacua are then represented by
\be
X_{2j}=\sqrt{\rho}, \qquad j=1,\ldots,M={N\over 2}
\ee
\begin{figure}
\bc
\epsfig{file=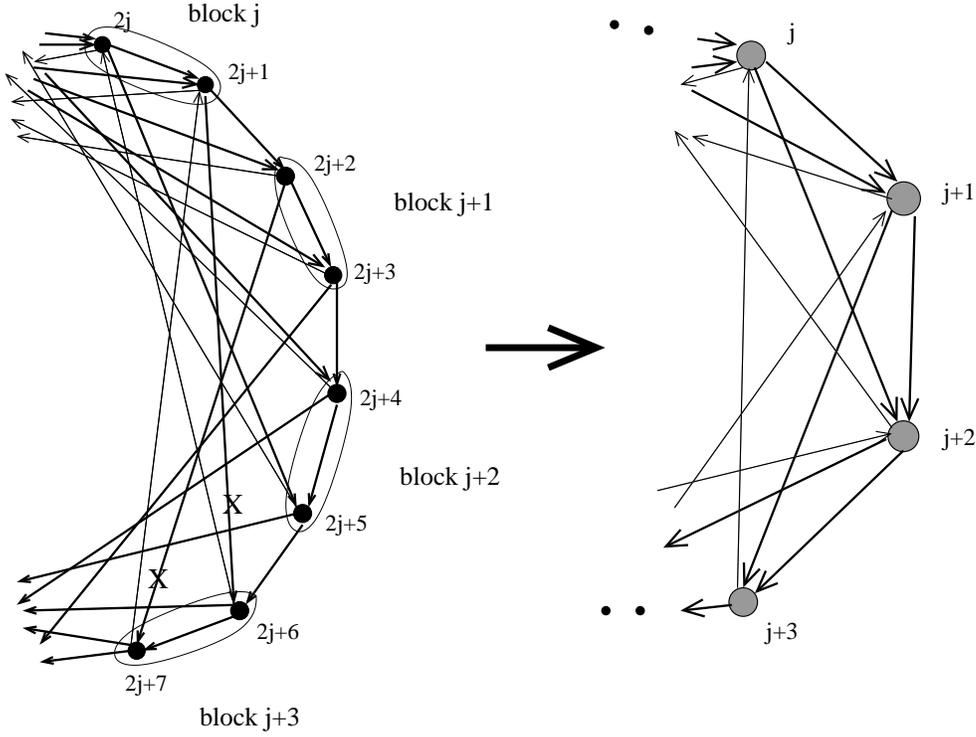, width=13cm}
\caption{Example $a=2$ : $\BC^3/\BZ_N : (1,5,-6) \ra (1,2,-3)$ : The links 
in the $(1,5,-6)$ quiver are $X_{i,i+1},Y_{i,i+5},Z_{i+6,i}$ and their 
conjugates. The process of coarse-graining as $(2j,2j+1)\ra j$ by 
condensing the $X_{2j,2j+1}$ link gives masses to the links with an X 
marked on them as well as the linear combinations $Z^+$. Retaining only 
the light links gives the quiver on the right, \ie\ $(1,2,-3)$.}
\label{fig6}
\ec
\end{figure}
as the only nonzero expectation values. This Higgses the $U(1)\times U(1)$
in each block down to the diagonal $U(1)$ as before. The superpotential is
\bea
&& W=W_2+W_3,\nonumber\\
&& {} 
W_2=\sum_{j=1}^M X_{2j} (Y_{2j+1} Z_{2j+2a+1} - Y_{2j-2a} Z_{2j+1})
\nonumber\\
&& {} \qquad
= \sqrt{\rho} \sum_j Z_{2j+1} (Y_{2j-2a+1} - Y_{2j-2a}) 
= \sqrt{\rho} \sum_j Z_{2j+2a+1} Y_j^-,
\nonumber\\
&& {} 
W_3=\sum_{j=1}^M X_{2j+1} (Y_{2j+2} Z_{2j+2a+2} - Y_{2j-2a+1} Z_{2j+2})
\eea
where we have defined $Y_j^{\pm}={1\over 2}(Y_{2j+1}\pm Y_{2j})$. Then
the $Z_{2j+1}$ and $Y_j^-$ fields appearing in $W_2$ become massive. The
classical equations of motion for $Z_{2j+1}$ following from $W$ give 
$Y_j^-=0$ so that
\be
Y_j^+={1\over 2}(Y_{2j}+Y_{2j+1})=Y_{2j}=Y_{2j+1}
\ee
Then the cubic part of the superpotential can be rewritten as
\bea
&& {} W_3=\sum_{j=1}^M X_{2j+1} [Z_{2j+2a+2} Y_{j+1}^+ - Z_{2j+2} 
Y_{j-a}^+]
=\sum_{j=1}^M Y_j^+ (X_{2j-1} Z_{2j+2a} - X_{2j+2a+1} Z_{2j+2a+2})
\nonumber\\
&& {}\qquad \equiv \sum_{j=1}^{M} 
Y_{j,j+a} (X_{j-1,j} Z_{j+a,j-1} - X_{j+a,j+a+1} Z_{j+a+1,j})
\eea
In the last line, we have as before relabelled $(2j,2j+1)\ra j$, so that 
this now looks like an \No\ superpotential for the light fields $X_{2j+1}, 
Y_j^+, Z_{2j}$ relabelled as $X_{j,j+1},Z_{j,j-a-1},Y_{j,j+a},\ 
j=1,\ldots,{N\over 2}$. The light link $Y_j^+$ in this case connects 
blocks $(j)$ and $(j+a)$ so that this again gives a $(1,a,-a-1)$ quiver. 
The residual quiver represents a $\BC^3/\BZ_{M}$ geometry with the orbifold
action $(1,a,-a-1)$ on the coordinates $X,Y^+,Z$. Figure~\ref{fig4} shows
how the $\BC^3/\BZ_{2M}\ (1,1,-2)$ quiver flows to the $\BC^2/\BZ_{M}$ 
quiver. Similarly the $\BC^3/\BZ_{2M}\ (1,2,-3)$ quiver flows to 
$\BC^3/\BZ_M\ (1,1,-2)$ (figure~\ref{fig5}). Figure~\ref{fig6} shows a 
slightly more complicated flow pattern. Thus in general, we think of the 
quiver as a lattice with $l$-th nearest neighbour interactions following 
from the orbifold projection. Then under the Higgsing, we flow 
progressively towards quivers with more local interactions.

\section{Coarse-graining the ``upstairs'' matrices}

In this section, we lift the Higgsing calculations described thus 
far to the ``upstairs'' $N\times N$ matrices of the image branes, thereby 
making connections to the matrix coarse-graining studied in \cite{kn0211}.
To this end, consider the scalars $X,Y,Z$ of the parent \Nf\ $U(N)$ theory 
after projection to the $\BC^2/\BZ_N$ orbifold, written as $N\times N$ matrices
\bea
&& {} X=\left( \bA{cccccc} 0 & X_{N1} & 0 & & & \\ 
& 0 & X_{12} & 0 & & \\ 
& & 0 & X_{23} & & \ldots \\
& & & 0 & & \\
& & & . & & \\
& & & . & & \\
\eA \right), \qquad
Y=\left( \bA{cccccc} 0 & & & & & Y_{N,N-1} \\ 
Y_{1N} & 0 & & & & \\ 
0 & Y_{21} & 0 & & & \\
& 0 & Y_{32} & \ldots & & \\
& & . & & & \\
& & . & & & \\
\eA \right),\nonumber\\
&& {} \qquad \qquad Z={\rm diag}[Z_{NN},Z_{11},Z_{22},\ldots,Z_{N-1,N-1}]
\eea
The orbifold projection ensures that these one-off-the-diagonal matrix 
elements are the only nonzero elements in these matrices. Further the 
projection also ensures that the gauge group is only the $U(1)^N$ subgroup
of the full $U(N)$ in the parent theory so that the notion of nearest 
neighbour is physically well-defined in these matrix representations.
These matrices do not commute for generic points on the orbifold moduli 
space, \ie\ generic link field expectation values. It is important to 
note that the status of these matrices for the image branes is different 
from those in \cite{kn0211}, since most of the entries do not exist here, 
due to the orbifold projection. We are simply using these ``upstairs'' 
matrices as a convenient device making contact with the ideas and 
calculations of \cite{kn0211}.

In Higgsing the $U(1)_{2j}\times U(1)_{2j+1}\ra U(1)$ in each pair (or
block) by condensing the links $X_{2j,2j+1}$, we see that we have
effectively bound the image branes as $(2j,2j+1)\ra j$ at lower
energies, thereby constructing a ``block-brane'' (for want of a more
apt word).  The fields $Y_{2j+1,2j},Z_j^-$ become massive.  In
\cite{kn0211}, subtracting off the diagonal modes after
coarse-graining in the (gauge-fixed) algebra of, \eg\ the noncommutative
plane, was found to yield a self-similar algebra with a new reduced
noncommutativity parameter.  Performing this procedure here produces the
${N\over 2}\times {N\over 2}$ matrices
\bea
&& {} {\tilde X}=\left( \bA{ccccccc} 0 & X_{12} & 0 & & & & \\ 
& 0 & X_{34} & 0 & & & \\ 
& & 0 & X_{56} & \ldots & & \\
& & & . & & & \\
& & & . & & & \\
\eA \right), \qquad
{\tilde Y}=\left( \bA{ccccccc} 0 & & & & & & Y_{N,N-1} \\ 
Y_{21} & 0 & & & & & \\ 
0 & Y_{43} & 0 & & & & \\
& 0 & Y_{65} & 0 & \ldots & & \\
& & & . & & & \\
& & & . & & & \\
\eA \right),\nonumber\\
&& {} \qquad \qquad
{\tilde Z}= {\rm diag}[{Z_{NN}+Z_{11}\over 2},{Z_{22}+Z_{33}\over 2}.
\ldots]
\eea
Note that the diagonal modes we have removed correspond precisely to
the fields $X_{2j,2j+1}$, $Y_{2j+1,2j}, Z_j^-$, which have been eaten
or given a mass by the Higgs mechanism. This is now the matrix
representation of the scalars on the partially resolved $\BC^2/\BZ_{N/2}$
orbifold that the ``block-brane'' lives on. This is an interesting
observation -- coarse-graining \cite{kn0211} of the scalar matrices
(with appropriate matrix-dependent averaging constants) followed by
subtracting the massive modes (which here are the diagonal elements
for the $X,Y$ matrices and the relative position coordinate in the $Z$
matrices) is precisely equivalent to the Higgsing analysis that we
have carried out in the previous sections. It is important to note
that the averaging constants are in general different for the
different scalar matrices : \eg\ $\al_X=\al_Y=1, \al_Z={1\over 2}$ is
the set of averaging constants for $\BC^2/\BZ_N$ above. This anisotropy
is not unexpected since the geometric meanings of the $X,Y$ matrices
as link fields and the $Z$ matrices as position coordinates are on
different footings. Furthermore the conditions in eqn.~(40) of
\cite{kn0211} are precisely the same as the hierarchies on the Higgs
expectation values for the $X_{i,i+1}$ with $\al_X=1$ 
\be 
\biggl|{{\tilde X}_{a,a+1}\over X_{a,a+1}}\biggr| 
= \biggl|{X_{2a,2a+1}\over X_{a,a+1}}\biggr| < 1 
\ee 
except that these conditions on the expectation values now
reflect the fact that the brane worldvolume gauge theory is
sequentially being truncated to lower energies under the process of
coarse-graining. It is amusing to note that these same equations
represent conditions for reducing noncommutativity (as defined in
\cite{kn0211}) from the point of view of the ``upstairs'' scalar
matrices. Furthermore, it is now clear that requiring these conditions
to be satisfied in the orbifold examples identifies precisely those
regions of moduli space that exhibit the corresponding hierarchies of
energy scales in the expectation values (or alternatively in the sizes 
of the closed string blowup modes in the geometry) ensuring that this
block-spin-like transformation is Wilsonian on the
worldvolume\footnote{Note that we have restricted attention to
homogeneous Higgsing in the link fields, engineering this in
accordance with homogeneous coarse-graining in the corresponding
matrices. However from the point of view of the orbifolds, we can
clearly turn on blow-up modes that are inhomogeneous in the quiver
lattice. Consider \eg\ $X_{2j}=v=\sqrt{\rho},\ j=l$ as the only
nonzero expectation value in $\BC^2/\BZ_N$, satisfying the F- and 
D-term equations with $D_{2l}=-D_{2l+1}=\rho$. This essentially 
``blocks up'' the $(2l)$-th and $(2l+1)$-th image branes, thus giving a 
``block-brane'' probe on a $\BC^2/\BZ_{N-1}$ orbifold. Note though that 
such inhomogeneous blocking on the quiver lattice essentially means 
that the corresponding matrix averaging as described above would be 
nonuniform, involving inhomogeneities in the matrix space.}.

It is easy to realize the same analysis for the other geometries we have
studied. For instance, consider the scalars $X,Y,Z$ of the parent \Nf\ 
$U(N)$ theory after projection to the $\BC^3/\BZ_N$ orbifold with action 
$(1,1,-2)$, written as $N\times N$ matrices
\bea
&& {} X=\left( \bA{ccccccc} 0 & X_{N1} & 0 & & & & \\ 
& 0 & X_{12} & 0 & & & \\ 
& & 0 & X_{23} & 0 & & \\
& & & 0 & X_{34} & \ldots & \\
& & & . & & & \\
& & & . & & & \\
\eA \right), \qquad
Y=\left( \bA{ccccccc} 0 & Y_{N1} & 0 & & & & \\ 
& 0 & Y_{12} & 0 & & & \\ 
& & 0 & Y_{23} & 0 & & \\
& & & 0 & Y_{34} & \ldots & \\
& & & . & & & \\
& & & . & & & \\
\eA \right),\nonumber\\
&& {} \qquad \qquad 
Z=\left( \bA{ccccccc} 0 & & & & \ldots & Z_{N,N-2} & 0 \\ 
0 & 0 & & & \ldots & 0 & Z_{1,N-1} \\
Z_{2N} & 0 & 0 & & & & \\ 
0 & Z_{31} & 0 & & & & \\
& 0 & Z_{42} & & \ldots & & \\
& & & . & & & \\
& & & . & & & \\
\eA \right)
\eea
As before, we coarse-grained the quiver as $(2j,2j+1)\ra j$, by 
condensing the links $X_{2j,2j+1}$ (these get eaten). Here it is the 
fields $Y_{2j-1,2j},Z_j^-$ that become massive. Then the above matrices 
coarse-grain as 
\bea
{\tilde X}=\left( \bA{ccccccc} 0 & X_{12} & 0 & & & & \\ 
& 0 & X_{34} & 0 & & & \\ 
& & 0 & X_{56} & & \ldots & \\
& & & . & & & \\
& & & . & & & \\
\eA \right), \qquad
{\tilde Y}=\left( \bA{ccccccc} Y_{1N} & 0 & & & & & \\ 
0 & Y_{23} & 0 & & & & \\
& 0 & Y_{45} & 0 & \ldots & & \\
& & & . & & & \\
& & & . & & & \\
\eA \right),\nonumber\\
{\tilde Z}=\left( \bA{ccccccc} 0 & & & & \ldots & Z_{N,N-2}+Z_{1,N-1} & \\ 
Z_{2N}+Z_{31} & 0 & & & & & \\ 
0 & Z_{42}+Z_{53} & 0 & & & & \\
& & & . & & & \\
& & & . & & & \\
\eA \right)
\eea
where we have retained only the light fields as before. Note that the 
scalar $Y$ has only diagonal elements now, so that it is a neutral field
now. This is the matrix representation of the scalars on the partially 
resolved $\BC^2/\BZ_{N/2}$ orbifold with coordinates $X,Z^+$ that the 
``block-brane'' lives on.\\
Consider now the scalars $X,Y,Z$ of the parent \Nf\ $U(N)$ theory after
projection to the $\BC^3/\BZ_N$ orbifold with action $(1,2,-3)$, written as 
$N\times N$ matrices
\bea
&& {} X=\left( \bA{ccccccc} 0 & X_{N1} & 0 & & & & \\ 
& 0 & X_{12} & 0 & & & \\ 
& & 0 & X_{23} & 0 & & \\
& & & 0 & X_{34} & \ldots & \\
& & & . & & & \\
& & & . & & & \\
\eA \right), \qquad
Y=\left( \bA{ccccccc} 0 & 0 & Y_{N2} & 0 & & \\ 
& 0 & 0 & Y_{13} & 0 & \\ 
& & 0 & 0 & Y_{24} & 0 & \\
& & & 0 & 0 & Y_{35} & \ldots \\
& & & . & & & \\
& & & . & & & \\
\eA \right),\nonumber\\
&& {} \qquad \qquad 
Z=\left( \bA{ccccccc} 0 & & & \ldots & Z_{N,N-3} & 0 & 0 \\ 
0 & & & \ldots & 0 & Z_{1,N-2} & 0 \\
0 & 0 & & \ldots & & 0 & Z_{2,N-1} \\
Z_{3N} & 0 & 0 & & & & \\ 
0 & Z_{41} & 0 & \ldots & & & \\
& & & . & & & \\
& & & . & & & \\
\eA \right)
\eea
We coarse-grained the quiver as $(2j,2j+1)\ra j$, by condensing the 
links $X_{2j,2j+1}$ (these get eaten). The fields $Z_{2j+3,2j},Y_j^-$ 
become massive. Then the above matrices coarse-grain as 
\bea
&& {} 
{\tilde X}=\left( \bA{ccccccc} 0 & X_{12} & 0 & & & & \\ 
& 0 & X_{34} & 0 & & & \\ 
& & 0 & X_{56} & \ldots & \\
& & & . & & & \\
& & & . & & & \\
\eA \right), \qquad
{\tilde Y}=\left( \bA{ccccccc} 0 & Y_{N2}+Y_{13} & 0 & & & & \\ 
& 0 & Y_{24}+Y_{35} & 0 & \ldots & & \\
& & & . & & & \\
& & & . & & & \\
\eA \right),\nonumber\\
&& {} {\tilde Z}=\left( \bA{ccccccc} 0 & & & & \ldots & Z_{N,N-3} & 0 \\ 
0 & 0 & & & \ldots & 0 & Z_{2,N-1} \\ 
Z_{41} & 0 & 0 & & \ldots & & \\ 
& & & & . & & \\
& & & & . & & \\
\eA \right)
\eea
where we have retained only the light fields. It is clear that this 
is the matrix representation of the scalars on the partially resolved 
$\BC^3/\BZ_{N/2}$ orbifold with action $(1,1,-2)$ on coordinates $X,Y^+,Z$ 
that the ``block-brane'' lives on.\\
It is not too hard to show that the matrix representations of the more 
general $\BC^3/\BZ_N$ quivers that we considered exhibit similar properties.
Note that in \cite{kn0211}, subtracting off the diagonal modes after 
coarse-graining in the (gauge-fixed) algebra of, \eg\ the noncommutative 
plane was found to yield a self-similar algebra with a new reduced 
noncommutativity parameter -- however it was not clear if this was a 
prescription compatible with Wilsonian renormalization organized by the
energy scales involved. We see now from the orbifold examples here that 
we can identify matrix coarse-graining as defined in \cite{kn0211} as a 
Wilsonian process if we first coarse-grain the matrices and then subtract 
off precisely the modes that are massive under the Higgs expectation 
values that we have given. It would be interesting to analyze more 
general D-brane configurations along these lines, incorporating the more 
serious issues of gauge invariance present in general.

It is an interesting question to generalize this equivalence between this
matrix coarse-graining, these block-spin-like transformations and field 
theoretic renormalization to more general scenarios with real D-branes 
and fluxes, with a view to understanding the flows in geometries that 
this generates.

\section{The structure of flows in the space of quivers}

To recap, in the region of moduli space in which we carried out our
calculations, the sequence of Higgs transitions represents a sequence
of partial blowups of the quotient singularity.  In the sense
discussed above in detail, it corresponds to block-spin-like
transformations on the image branes.  Thinking of the quiver as a
lattice these were analogous to coarse-graining
transformations. Furthermore, lifting this process to the ``upstairs''
matrices makes contact with the matrix coarse-graining of
\cite{kn0211}. This process can be represented diagrammatically as we
have seen in the figures. In this section, we will map out the structure
of the flows we have obtained.

Under a $(2\times 2)$-block coarse-graining, we have seen that, \eg\ a
$\BC^3/\BZ_N\ (1,2a,-2a-1)$ quiver flows to a $\BC^3/\BZ_{N/2}\
(1,a,-a-1)$ quiver. Continuing the process we see from the change in
$a$ that the quiver is becoming progressively ``more local.''
Eventually, the process leads to a $\BC^2/\BZ_{N'}$ quiver.  In our
spin system analogy, this type of quiver corresponds to a local
interaction (with range small compared to the correlation length of
the system).  By contrast, if we consider the quotient with $a$ of
order $N$, which corresponds to a highly nonlocal interaction, our
successive coarse-graining transformations produce a sequence of
quivers with comparable measures of ``non-locality.'' Further, while
we have not shown this explicitly, it is not too hard to see from the
quiver diagrams that changing block size to, say $l\times l$ blocks,
does not change the qualitative structure of the coarse-graining we
have shown for these quivers.

\begin{figure}
\bc
\epsfig{file=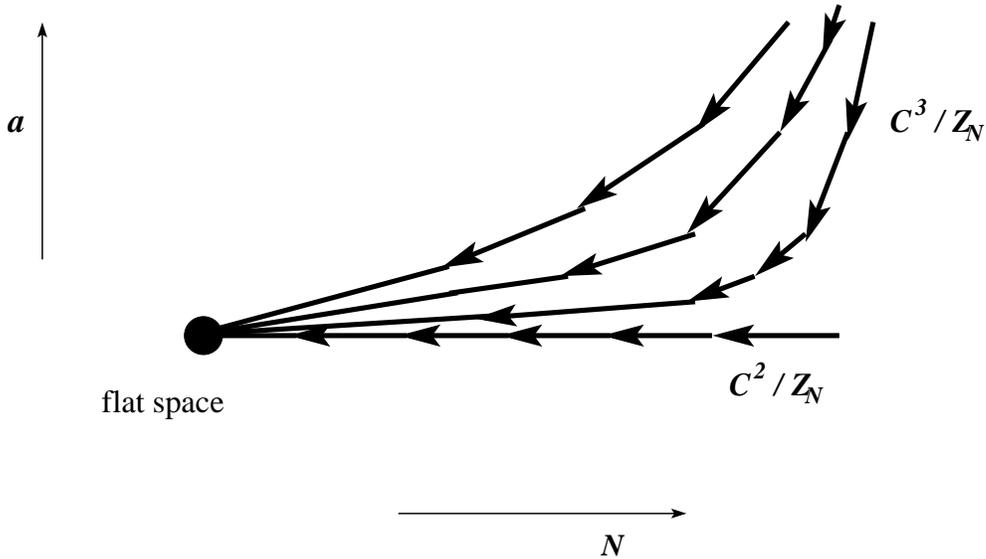, width=13cm}
\caption{The structure of flows : The general point in this simplified 
subset of the space of quivers is $\BC^3/\BZ_N,\ (1,a,-a-1)$, which we
have labelled in this $(N,a)$-plane. Flat space with \Nf\ supersymmetry
is a fixed point. All flowlines end finally at flat space. However the
$\BC^3/\BZ_N,\ a \ll N$ quivers flow towards $\BC^2/\BZ_N$ quivers before
flowing finally to flat space, enhancing supersymmetry from \No\ to \Nt\
to finally \Nf.}
\label{fig7}
\ec
\end{figure}

Consider now the space of quiver models, restricting attention to
lowest order in $\al'$ and $g_s$, with no background fluxes. A point
in this subset can be described as $\BC^3/\BZ_N$ with orbifold action
$(1,a_2,a_3)$. This simplified subset of the space of quiver models is
thus a lattice-like space: a point in this space is labelled by
$(N,a_2,a_3)$. Restricting attention to supersymmetric vacua is
equivalent to restricting to the subspace with $\ 1+a_2+a_3=0$.  Under
these block-spin-like transformations, we have seen that $\BC^3/\BZ_N$
for finite $N$ flows to the point $(0,0,0)$: this is flat space and it
is a fixed point. The number of iterations required is $O(\log
N)$. Thus as $N$ grows large, the number of iterations diverges and an
RG-like approach is valid.  We have also seen that the $\BC^2/\BZ_N$
quiver for any $N$ flows towards flat space. Furthermore, the
$\BC^3/\BZ_N,\ a_i \ll N$, quivers flow towards $\BC^2/\BZ_N$ quivers,
before finally flowing to flat space, enhancing supersymmetry from \No\ 
to \Nt\ to finally \Nf. Figure~\ref{fig7} summarizes the flows in quiver
space.

It is worthwhile to note that these theories that we have been 
considering are each good supersymmetric D-brane configurations (or 
string vacua) in themselves. Thus this block-spin-like transformation we 
have described here is really a flow in the (simplified subset of the)
space of D-brane configurations. 

Now it is clear that if the brane sits $at$ the orbifold singularity,
all the link modes are massless and there is no hierarchy of scales.
However in the region of moduli space where the link expectation values 
exhibit a wide hierarchy of scales, this transformation exhibits an 
interesting structure of flows and fixed points that are self-similar. 
Note though that this region of moduli space is not unique : \eg\ for 
$\BC^2/\BZ_N$ via $(2\times 2$ blocks, we condensed links $X_{2j,2j+1}$, 
which is Wilsonian in the region with the hierarchies $|X_{12}|<|X_{N1}|$ 
and so on, while condensing the $X_{2j-1,2j}$ links gives a different 
region of moduli space. However while the regions in moduli space are 
different, the qualitative features of the flows -- the directions of 
the flows and the fixed points -- are the same. 

The description above is of course a highly simplified picture of the 
space of quiver gauge theories appearing on the worldvolumes of D-branes
probing orbifolds. Considering more general theories (\eg, $\BR^6/\Gamma$,
where $\Gamma$ is some discrete group, possibly with additional 
background fluxes turned on) imparts further structure to this space of 
quivers. It would be interesting to understand if this or any other kind 
of RG-like transformation can be constructed in more generality, possibly 
relating it to the ideas of \cite{vacuastats}.

\section{Conclusions}

We have described a block-spin-like transformation that coarse-grains 
quivers by sequentially Higgsing the gauge symmetry in these theories 
and exhibited the pattern of flows obtained thereby.  Since the points
in the space of quivers that we have discussed above are distinct string 
vacua, this transformation generates flows in this simplified subset of 
the space of D-brane configurations (or string vacua). In hindsight, the 
general schematic picture of matrix coarse-graining described in 
\cite{kn0211} would heuristically be expected to generate flows in the
space of D-brane configurations, modulo the gauge-fixing problems there.
It is thus an interesting open question to extend this block-spin-type 
procedure we have described here to more general D-brane configurations. 
Unlike the quiver examples here, understanding the analog of 
``nearest neighbour'' is more subtle in general, \eg\ where the gauge 
group is $U(N)$. We expect that this question is closely intertwined 
with issues of locality in both real space and its reflection in the 
space of matrices, as noted in the gauge-fixing problems observed in 
\cite{kn0211}. Understanding how to incorporate locality in a 
gauge-invariant fashion might in turn be related to hierarchies in the 
strengths of the link expectation values, something that has been 
relatively easy to nail down in the quiver examples here. It is 
interesting to ask if there is any notion of universality in the flows 
that would be generated in the space of geometries for general D-brane 
configurations, or more generally string vacua.

As we have seen in section 4, our description of the quiver flows makes 
contact with the matrix coarse-graining of \cite{kn0211} when the 
Higgsing process is lifted to the ``upstairs'' $N\times N$ matrices of 
the $N$ image branes for a $Z_N$ quotient singularity. These ``upstairs''
matrices are noncommuting for generic expectation values of the 
bifundamental fields ``downstairs'' (\ie\ generic points in the orbifold 
moduli space). Then, as have seen in section 4, the conditions for 
reducing noncommutativity (as defined in \cite{kn0211}) of the 
``upstairs'' matrices coincide with the conditions on the region of 
moduli space for the process to be Wilsonian on the worldvolume 
``downstairs''. Thus the orbifold computations described here provide a 
concrete field theoretic realization of the matrix coarse-graining of 
\cite{kn0211}. Furthermore, since the gauge group after orbifold 
projection is $U(1)^N$, the gauge invariance problems of the matrix RG 
of \cite{kn0211} are absent here. 

For the quiver theories, this process defines a kind of deconstructive 
RG in theory space with the corresponding RG flows as described in the 
previous section. This is a kind of real space RG that coarse-grains the 
extra dimension. From this point of view, we find that the large $N$ 
limit of $\BC^2/\BZ_N$ (in this region of orbifold moduli space) appears 
to deconstruct an inhomogeneous continuum extra dimension. Such an 
extra dimension is fractal in the sense that there is self-similarity at
arbitrarily short distances. Note that this process is different from 
applying conventional blockspin-type techniques on the lattice gauge 
theory in the extra dimension\footnote{See \eg\ \cite{hillDeconsRG} for a 
different discussion of fractal theory space and RG therein. See also 
\cite{deconsTop} for discussions on the topology of theory space.}. A 
uniform extra dimension, generated by equal expectation values for the 
link fields, does not exhibit any hierarchy of energy scales and so an 
RG by sequentially integrating out massive link expectation values is not 
possible along the lines we have studied here. Presumably conventional 
blockspin-type techniques (or momentum space RG in the extra dimensions) 
could be applied to study such an RG at a field theoretic level.

Finally, it is important to note that this scheme of coarse-graining via 
condensing link expectation values in these supersymmetric theories turns 
on marginal deformations from the point of view of the corresponding 
worldsheet string theory in this background. A natural question is what 
the corresponding story is for nonsupersymmetric orbifolds with localized 
closed string tachyons \cite{aps} \cite{hkkm}. These orbifolds typically 
have both relevant (tachyonic) and marginal closed string twisted sector 
modes. Assuming that the system does not run away along the directions 
with tachyonic instabilities, we can extend the above analysis to the 
marginal modes. On the other hand, turning on a tachyonic (relevant) 
deformation on the worldsheet corresponds to evolution in time in the 
target space, \ie\ the worldsheet RG scale corresponds to time in the 
target space. It is interesting to ask what the relation of this 
block-spin-like transformation is, if any, to these tachyonic flows.

\vspace{6mm}

{\small {\bf Acknowledgments:} It is a pleasure to thank P.~Argyres,
P.~Aspinwall, M.~Douglas, P.~Horava, S.~Kachru, I.~Melnikov, D.~Morrison,
M.~Rangamani, S.~Rinke, A.~Sen, S.~Trivedi, M.~van Raamsdonk and
H.~Verlinde for helpful discussions. KN thanks the organizers of the
HRI String workshop, Allahabad, India, for hospitality during incipient 
stages of this work, as well as hospitality of the Theory Group
at UC Berkeley during stages of this work. MRP thanks KITP and the 
organisers of the Geometry, Topology, and Strings workshop (MP03) where 
some of this work was completed.  His participation was supported in 
part by the National Science Foundation under Grant No. PHY99-07949.  
This work is partially supported by NSF grant DMS-0074072.}

\appendix
\section{Interpolating expectation values}

In this appendix, we construct a set of Higgs expectation values which 
interpolate between the equal expectation values limit $h\neq 0, 
\rho_i\sim 0$ and our hierarchical limit $h\sim 0, \rho_i\neq 0$. Let 
us do this sequentially so as to break the $Z_N$ discrete symmetry but 
retain a $Z_{N/2}$ subgroup thereof. If we restrict attention to real 
expectation values for $X_i, Y_i$ with $Z_i=0$, it is sufficient, 
instead of (\ref{NtFterms}) and (\ref{NtDterms}), to solve 
\bea
X_{2j}^2Y_{2j}^2=X_{2j-1}^2Y_{2j-1}^2=X_{2j-2}^2Y_{2j-2}^2, && {}
\nonumber\\
|X_{2j}|^2-|Y_{2j}|^2-|X_{2j-1}|^2+|Y_{2j-1}|^2=\rho_{2j}&=&\rho,
\nonumber\\
|X_{2j-1}|^2-|Y_{2j-1}|^2-|X_{2j-2}|^2+|Y_{2j-2}|^2=\rho_{2j-1}&=&-\rho
\eea
to execute sequential Higgsing. It is clear that a solution that preserves 
the $Z_{N/2}$ subgroup can be found by redefining
\be
a_{2j-1}=X_{2j-1}^2=a_1, \ b_{2j-1}=Y_{2j-1}^2=b_1, \ \ \
a_{2j}=X_{2j}^2=a_2, \ a_{2j}=Y_{2j}^2=b_2, \qquad j=1,\ldots {N\over 2}
\ee
to find a solution to this $2\times 2$ block within the $Z_N$ and then
repeat it $N/2$ times preserving $Z_{N/2}$. Thus it is sufficient to find
solutions to
\be
a_2 b_2 = a_1 b_1, \qquad a_2 - a_1 - b_2 + b_1 = \rho
\ee
such that for ${\rho\over h}\ll 1$, we have the probe on the unresolved
orbifold with $a_2=a_1=b_1=b_2=h$, while ${\rho\over h}\gg 1$ yields 
the partially resolved limit with $a_2=\rho,a_1=b_1=b_2=0$. It is not too
hard to show that
\bea
x=1-\tanh {\rho\over h}, \qquad \qquad X_{2j-1}^2=a_1=h x, && {} \qquad 
X_{2j}^2=a_2=h\biggl({{\rho\over h}+x+x^2 \over 1+x}\biggr), \nonumber\\
Y_{2j-1}^2=b_1=h x \biggl({{\rho\over h}+x+x^2 \over 1+x}\biggr),
&& {} \qquad Y_{2j}^2=b_2=h x^2
\eea
is a set of solutions to the F- and D-term equations that does the 
interpolation. This gives rise to a hierarchy of scales where the $X_{2j}$
expectation values become more massive than the rest as we increase $\rho$.

\end{document}